\documentclass[preprint,preprintnumbers,amsmath,amssymb,nofootinbib,superscriptaddress]{revtex4}

\usepackage{graphicx}% Include figure files
\usepackage{dcolumn}% Align table columns on decimal point
\usepackage{bm}% bold math

\begin{document}

%\preprint{}

\title{ Critical magnetic field in AdS/CFT superconductor}

\author{Eiji Nakano}
\email{e.nakano@gsi.de}
\affiliation{Department of Physics and Center for Theoretical Sciences\\ 
National Taiwan University, Taipei 106, Taiwan}
\affiliation{Gesellschaft f\"{u}r Schwerionenforschung, GSI, D-64291 Darmstadt, Germany}
\author{Wen-Yu Wen}
\email{steve.wen@gmail.com}
\affiliation{Department of Physics and Center for Theoretical Sciences\\ 
National Taiwan University, Taipei 106, Taiwan}

%\date{\today}

\begin{abstract}
We have studied a holographically dual description of superconductor 
in $(2+1)$-dimensions in the presence of applied magnetic field, and 
observed that there exists a critical value of magnetic field, 
below which a charged condensate can form via a second order phase transition. 
\end{abstract}

%\keywords{}

\maketitle

\section{Introduction}
The holographic correspondence between a gravitational theory and a quantum field theory, first emerged under the AdS/CFT correspondence\cite{Maldacena:1997re}, has been proved useful to study various aspects of nuclear physics such as RHIC and condensed matter phenomena, particularly in those recent studies \cite{Herzog:2007ij,Hartnoll:2007ih,Hartnoll:2007ip,Hartnoll:2008hs,Minic:2008an}.

In the papers \cite{Gubser:2005ih,Gubser:2008px}, the author proposed a gravity model in which Abelian symmetry of Higgs is spontaneously broken by the existence of black hole.  This mechanism was recently incorporated in the model of superconductivity and critical temperature was observed\cite{Hartnoll:2008vx}, and later on non-Abelian gauge condensate\cite{Gubser:2008zu}.  In this paper, we would like to extend the work to include the magnetic field and will show the existence of critical magnetic field as expected from physics of superconductor.

To implement a magnetic field at finite temperature, we introduce a Reissner-Nordstrom charged black hole and a condensate through a charged scalar field.  In the superconducting phase, the scalar field takes different values at the horizon for different condensate expectation value at the boundary, indicating the existence of a scalar hair; while in the normal phase, vanishing scalar field tells the ordinary tale of a black hole with no hair.

\section{The model with applied magnetic field}
Several important unconventional superconductors, such as the cuprates and organics, are layered in structure and interesting physics can be captured by studying a $(2+1)$ dimensional system.  We are now interested in building up a gravity model (in coupled with other matter fields) in $(3+1)$ dimensions which is holographically dual to the desired planar system which develops superconductivity below critical temperature and critical magnetic field.  We start with a model composed of the gravity sector and the matter sector.  The gravity sector is given by the following Lagrangian density,
\begin{equation}
e^{-1}{\cal L}_g=R-\frac{6}{L^2}-\frac{1}{4}{\cal F}^{\mu\nu}{\cal F}_{\mu\nu},
\end{equation}
together with a solution of {\sl magnetically} charged black hole in $AdS_4$, where\cite{Romans:1991nq} 
\begin{eqnarray}
&&ds^2=-f(r)dt^2+\frac{dr^2}{f(r)}+r^2(dx^2+dy^2),\\\nonumber\\
&&f(r)=\frac{r^2}{L^2}-\frac{M}{r}+\frac{H^2}{r^2}.
\end{eqnarray}
Through the paper we set radius of curvature $L=1$ for numerical computation.  By assumption the only nonzero electro-magnetic field is the magnetic component ${\cal F}_{xy}=\frac{H}{r^2}$, of which the energy density at any fixed radius coordinate $r$ is always finite and constant, that is, ${\cal F}^{\mu\nu}{\cal F}_{\mu\nu}\propto H^2$.  This serves the purpose of constant applied magnetic field at the boundary.  The black hole is censored by a horizon provided the condition $27M^4-256H^6\ge 0$ and the temperature, as a function of $M$ and $H$, is determined via the relation 
\begin{equation}
T=\frac{f'(r_+)}{4\pi},
\end{equation}%The extremal limit happens at $f(r)=f'(r)=0$, that is $r_+=r_-=(\frac{H^2}{3})^{1/4}=(\frac{M}{4})^{1/3}$.
where $r_+$ is the most positive root of $f(r)=0$ (outer horizon).  We expect that the gravity sector, implied by its given name, can be easily obtained from a pure gravity theory of higher dimensions by appropriate reduction.

For the matter sector, we will use the Ginzburg-Landau (GL) action for a Maxwell field and a charged complex scalar, which does not back react on the metric \cite{Gubser:2008px,Hartnoll:2008vx}, 
\begin{equation}
e^{-1}{\cal L}_m=-\frac{1}{4}F^{ab}F_{ab}+\frac{2}{L^2}|\Psi|^2-|\partial \Psi-iA\Psi|^2.
\end{equation}
This action differs from the usual GL theory by two places: the coefficient of $|\Psi|^2$ term appears to be negative in both ordinary and superconducting phase, and a $|\Psi|^4$ term is not included.  The AdS bulk geometry, however, plays the role of stabilization and we still expect some kind of Higgs mechanism triggered outside the horizon\cite{Gubser:2008px}. Enough for our purpose, we will also assume the planar symmetry ansatz for the scalar potential $A_t=\Phi(r)$ and the complex scalar $\Psi(r)$, where we have already fixed the phase to be constant.  Then we need to solve a pair of coupled second order differential equations
\begin{eqnarray}\label{eqn:2ode}
&&\Psi''+(\frac{f'}{f}+\frac{2}{r})\Psi'+\frac{\Phi^2}{f^2}\Psi+\frac{2}{L^2f}\Psi=0,\nonumber\\
&&\Phi''+\frac{2}{r}\Phi'-\frac{2\Psi^2}{f}\Phi=0
\end{eqnarray} 
with appropriate boundary conditions at the horizon and at asymptotic infinity.  They can be solved numerically regardless of  difficulty which appears in finding nontrivial analytic solutions.  In particular, for normalizable scalar potential, we require at the horizon\cite{Gubser:2008px,Hartnoll:2008vx}
\begin{eqnarray}
&&\frac{\Psi'}{\Psi}\bigg|_{r=r_+}=\frac{-2r_+}{3r_+^2-\frac{H^2}{r_+^2}},\nonumber\\
&&\Phi(r_+)=0.
\end{eqnarray}

%%%%%%%%%%%%%%%%%%%%%%%%%%%%%%%%%%%%%%%%%%
\begin{figure}\label{fig_1}
\includegraphics[width=0.45\textwidth]{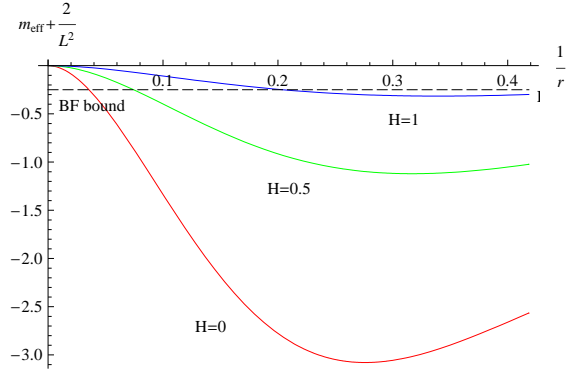}
\caption{The effective mass $m_{eff}^2$ evaluated at fixed temperature and boundary conditions at the horizon.  From bottom up, the curves are with $H=0,0.5$ and $1$ respectively.  The dashing line indicates the Bretenlohner-Freedman bound, below which the AdS vacuum is unstable under perturbation of $\Psi$ and condensation is expected.}
\end{figure}
%%%%%%%%%%%%%%%%%%%%%%%%%%%%%%%%%%%%%%%%%

Nevertheless we still have freedom for two parameter family of solutions by assigning $\Phi'$ and $\Psi$ at the horizon, therefore we have a scalar hair from black hole for non-vanishing $\Psi$.
At the boundary, the solutions behave like
\begin{eqnarray}
&&\Psi=\frac{\Psi_{(1)}}{r}+\frac{\Psi_{(2)}}{r^2}+\cdots,\nonumber\\
&&\Phi=\mu-\frac{\rho}{r}+\cdots,
\end{eqnarray}
where $\mu$ and $\rho$ are interpreted as chemical potential and 
charge density in the dual field theory.  
We are interested in the case where either $\Psi_{(1)}$ or $\Psi_{(2)}$ 
vanishes for stability concern at asymptotic AdS region, 
then read off the pairing operator $\cal O$ dual to $\Psi$ 
from the bulk-boundary coupling \cite{Hartnoll:2008vx}, 
\begin{equation}
\langle{\cal O}_i\rangle=\sqrt{2}\Psi_{(i)}. 
\end{equation}

To gain a better intuition of how a condensate is realized 
in this gravity setup, 
we may investigate the effective mass of $\Psi$ field along 
the radius direction, that is
\begin{equation}
m^2_{eff}(r)=-\frac{2}{L^2}-\frac{\Phi^2}{f}.
\end{equation}
We recall that there exists the Bretenlohner-Freedman (BF) bound\cite{Breitenlohner:1982bm}, i.e. $m^2L^2>-9/4$, which guarantees that the AdS vacuum is stable under perturbations of $\Psi$.  We observe that in the Figure 1 that provided fixed temperature and boundary condition at the horizon, the effective mass dives below the BF bound for wider range of $r$ for smaller magnetic field.  In the other words, condensate happens more easily while the magnetic field is smaller.  This implies the existence of critical magnetic field below which the condensation can take place.

%%%%%%%%%%%%%%%%%%%%%%%%%%%%%%%%%%%%%%%%%%%%%%%%%%%%%%%%%%%%
\begin{figure}\label{fig_2}
\includegraphics[width=0.45\textwidth]{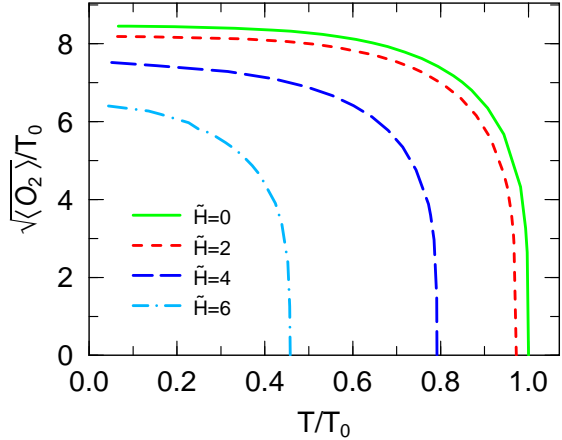}
\caption{We plot order parameter $\langle{\cal O}_2\rangle$ 
as a function of temperature.  The critical temperature $T_c$ decreases as applied magnetic field increases.  Here $\tilde{H}$ is the normalized $H$ given by $H^{2/3}/T_0$, where $T_0=T_c$ at $H=0$.}
\end{figure}
%%%%%%%%%%%%%%%%%%%%%%%%%%%%%%%%%%%%%%%%%%%%%%%%%%%%%%%%%%%%

\section{critical magnetic field}
In the normal phase, we always have solutions to 
the equations (\ref{eqn:2ode}), that is $\Psi=0$ 
and $\Phi=\mu-\frac{\rho}{r}$; 
while in the superconducting phase, 
we may have nontrivial $\Psi(r)$ and 
its boundary value serves as an order parameter for condensate.  
In the absence of applied magnetic field, for any fixed $\rho$, 
there exists a critical temperature $T_c$, 
above which there is no more nontrivial solution\cite{Hartnoll:2008vx}.  In the presence of applied magnetic field, however, the Meissner effect is expected and there exists both $T_c$ and a critical magnetic field $H_c$, above which the nontrivial solution is again not admissible.  As argued in the previous section, we expect that the stronger applied magnetic field $H$ is, the lower is critical temperature $T_c$.  
This statement is supported by our numerical 
results for $\langle{\cal O}_2\rangle$ 
as shown in the Figure 2. 
The operator ${\cal O}_{2}$ corresponds to a pair of fermions, 
while ${\cal O}_{1}$ to a pair of bosons\cite{Hartnoll:2008vx}. 
We have also found similar results for $\langle{\cal O}_1\rangle$ 
only at a small $H$ region.

In the Figure 3 we also plot the phase diagram of critical magnetic field 
against critical temperature.   

%%%%%%%%%%%%%%%%%%%%%%%%%%%%%%%%%%%%%%%%%%%%%%%%%%
\begin{figure}\label{fig_3}
\includegraphics[width=0.45\textwidth]{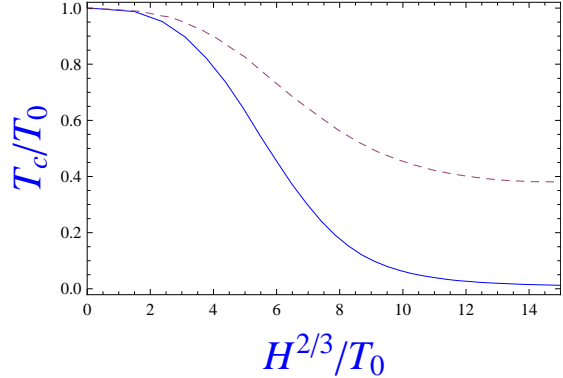}
\caption{The phase diagram of $T_c$ against $H_c$.  
The superconducting phase where $\langle{\cal O}_2\rangle \neq 0$($\langle{\cal O}_1\rangle \neq 0$)
exists in the lower left part below the solid (dashed) curve, 
while normal phase in the upper right part above the curve.  
}
\end{figure}
%%%%%%%%%%%%%%%%%%%%%%%%%%%%%%%%%%%%%%%%%%%%%%%%%

\section{Discussion}
In this paper, 
we have considered a hybrid model for AdS/CFT superconductors 
in the presence of magnetic field.  Several comments are in order: 
At first, a magnetic field is provided in the gravity sector 
as a background, independent of the probed sector.  
We argue that this is perfectly fine 
as long as we only consider a constant magnetic field at the boundary. 
Secondly, the matter sector has no back reaction to the gravity sector, 
therefore the equation of motion for total Lagrangian is not satisfied.  
Although this may not be crucial to the occurrence of 
superconducting phase, 
it is still interesting to investigate a fully back-reacted action 
which can be derived from some higher-dimensional theory 
such as String theory or M-theory.
Thirdly, in order to discuss possible formation of vortex lattice 
and distinguish between type I and II superconductors, 
one is tempted to relax the ansatz of planar symmetry.  
This will complicate the construction and analysis and we hope 
to report it in the near future.  At last, 
this construction is a tractable model of strongly coupled system 
which may capture some physics of unconventional superconductors, 
in contrast to the conventional superconductors well 
described by GL theory macroscopically and BCS theory microscopically.  
Though we do not see fermionic degree of freedom from 
this macroscopic construction, the complex scalar, 
serving as order parameter, 
seems sufficient to explain such a critical phenomenon 
as good as the usual GL theory.  
In order to pursue a microscopic model along this line of reasoning, 
one may still need to understand better how to realize underlying 
fermionic degree of freedom in the context of AdS/CFT correspondence.

\begin{acknowledgments}
The authors are partially supported by 
the Taiwan's National Science Council and National Center 
for Theoretical Sciences under Grant No. NSC96-2811-M-002-018, 
NSC97-2119-M-002-001, and NSC96-2811-M-002-024.
\end{acknowledgments}

%%%%%%%%%%%%%%%%%%%%%%%%%%%%%%%%%%%%%%%%%%%%%%%%%%%%%%%%%%%%%%%

\bibliography{apssamp}% Produces the bibliography via BibTeX.

\end{document}